\documentstyle[12pt]{article}
\title{
Calorimetric tunneling study of heat generation\\
in metal-vacuum-metal tunnel junction
}
\author{Ivan Bat$\!$'ko and Marianna Bat$\!$'kov\'a\\
Institute  of  Experimental  Physics, 
 Slovak   Academy   of Sciences\\
Watsonova 47, 043 53 Ko\v {s}ice, Slovakia\\
\\}
\date{}

\begin{document}
\baselineskip=14pt
\maketitle

\begin{abstract}
We have proposed novel calorimetric tunneling (CT) experiment
 allowing exact determination of heat generation  (or heat sinking) in individual
 tunnel junction (TJ) electrodes which opens new possibilities in the field of design
 and development of experimental techniques for science and technology.
    Using such experiment we have studied the process of heat generation in normal-metal electrodes
of the vacuum-barrier tunnel junction (VBTJ).
The results show there exists dependence
of the mutual redistribution of the heat on applied bias voltage
and the direction of tunnel current, although the total heat generated
in tunnel process is equal to Joule heat, as expected.
    Moreover, presented study indicates generated heat represents
the energy of non-equilibrium quasiparticles coming from inelastic electron
processes accompanying the process of elastic tunneling.
\end{abstract}
\thanks{PACS nrs.:73.40.Gk,  73.40.RW, 72.15.Jf, 07.79.-v}

\newpage

     A deeper understanding of the power dissipation
 due to charge injection in tunnel junctions (TJs)
 and point contacts (PCs) is of a great
importance for optimal design, fabrication and further
miniaturization of nano-scale electronic devices.
    One of the important points is the detail knowledge of the nature
of the heat generation and an appropriate characterization of
energy dissipation processes in such structures.
     Experiments with
ballistic PCs \cite{PCS} show asymmetry of heat generation
associated with the fact that the charge carriers accelerated in
the field region of the contact propagate ballistically, i.e.
without scattering, and dissipate gained energy generating
non-equilibrium quasiparticles only {\em after} passing the contact
\cite{Ursula83,Reiffers86}.
    This asymmetry can be even a source of information
about  electron-phonon interaction \cite{Reiffers86,Bogacek85}.
  Observations of dissipation asymmetries in TJs
at high bias \cite{Hebard76} and studies of self-heating due to
electron tunneling in the Coulomb-blockade electrometer
\cite{Kautz92}, as well as investigations of electronic
refrigeration in the normal-metal -- insulator -- superconductor
(NIS) TJ \cite{Nahum94} show that  the process of charge tunneling
is frequently associated with generation of thermal gradients.
  This implicate the importance of experimental studies
yielding the information about the heat generation or
heat sinking in individual TJ electrodes.

Analyzing properties of the tip-sample configuration as well-known
from scanning tunneling microscopy (STM), we have formed an opinion that
utilization of the VBTJ should
allow an exact calorimetric determination (measured in absolute units)
of the heat power dissipated in each of TJ electrodes
as the vacuum tunnel barrier secures a thermal decoupling of the electrodes.
    Moreover, unlike experiments
where TJ is placed e.g. in liquid helium \cite{Hebard76},
the studies performed at vacuum conditions remove problems coming from
parasitic thermal coupling of TJ electrodes with their
surrounding.

   In this paper  we explain the principle of
the proposed CT experiment
and present results of exact calorimetric measurements of the heat generation in the
{\em separate} electrodes of normal-metal -- vacuum -- normal-metal TJ
obtained by this novel experimental technique.
 Our results show that the mutual redistribution of the heat power
dissipated in (normal) metal electrodes of TJ depends on bias
voltage applied over the TJ and it can be adequately described within
the generally accepted conception of electron tunneling
\cite{Hebard76,Kautz92,Wolf85}.

  In principle, CT experimental setup
consists of the STM-like  tunneling unit and sample holding calorimeter
 for specific heat measurements by relaxation method,
e.g. like used in \cite{Schutz74,Meissner}.
  The experimental setup used for studies presented in this paper
is depicted in Fig.~1.
    The tip is  connected to the linear piezo-positioner \cite{LAR}
 of the low temperature high vacuum compatible tunneling head
and its separation from the sample fixed to the sapphire plate is controlled
by  the z-feedback STM control electronics \cite{LAR}.
    The sapphire plate equipped by a bare chip Ge-thermometer
and a RuO$_2$ resistor as a heater is supported from one side
by three stainless steel needles fixed to the
copper block and from the opposite site it is
pressed by warped phosphor bronze needle
thermally linked to the same copper block.
   All, the sample, the thermometer and the  heater are glued to
the plate using a small amount of GE-7031 varnish.
     Such design, similar to one of calorimeter for
specific heat studies by relaxation method \cite{Meissner}
and calorimetric absorbtion spectroscopy studies \cite{CAS}
has shown a sufficient mechanical stability and
a plausible heat resistance for CT experiments
at temperatures up to 6~K at least.

         Both, the tip and the sample
were cut from the  same gold
based alloy Au$_{0.7}$Cu$_{0.16}$Ag$_{0.14}$
in order to prevent a formation of insulating layer
on the tip and the sample surfaces and to avoid effects due to
dissimilarities of TJ electrodes.
     Immediately after cleaning of the tip and the sample surfaces
the experiment was placed into the  vacuum space of $^4$He cryostat,
 pumped to the
high vacuum pressure at simultaneous  overheating to the temperature
above 40~\raisebox{1ex}{\scriptsize o}C and then slowly cooled
down to low temperatures.

   The data were taken at temperatures close to 5.3~K.
 The heat power generated in the sample due to the tunnel current $I$
was derived from the increase of the sapphire plate temperature.
    The experimental procedure consisted of
the determination of the generated heat power $P_{pos}$ and $P_{neg}$
for positively and negatively biased sample, respectively,
at the same absolute value of tunnel current $I$,
with the aim to get
an  {\em exact comparison}
of the heat power generated in the sample
due to electron tunneling for ``direct" and ``reverse"
polarity of the tunnel current.

 	Measurements at constant absolute value of $I$ (Fig. 2)
 show that in the low-voltage limit ($V<300$~mV)
  $P_{pos}$ and $P_{neg}$ are equal within the resolution
of experiment.
At higher bias voltage $V$ the heat power generated in the sample
shows a clear  asymmetry
with respect to the orientation of $I$.
The data show that
$P_{pos}$ (electrons injected {\em into the sample})
is greater than $P_{neg}$ (electrons injected {\em from the sample}).
    The difference between  $P_{pos}$ and $P_{neg}$
{\em nonlinearly} increases with increasing $V$.
    On the other hand,
    $P$-$I$ dependencies
at constant absolute value of $V$ (inset of Fig. 2.) show
that
the  ratio $P_{pos} / P_{neg}$ does not depend on $I$
as both, $P_{pos}$ and $P_{neg}$
 are linear functions of $I$
within the resolution of experiment and scanned range of parameters.
  Because of the same material of the tip
 and the sample, no effects  due to dissimilarities
of the TJ electrodes
are considered and the total power   generated in the VBTJ
can be expressed as  $P_{tot}= P_{pos} + P_{neg}$.
    As shown in Fig. 2, including its inset,
there is an excellent coincidence
of $P_{tot}$
 with  Joule heat power $P_{Joule} =VI$
 calculated for electrically equivalent
 resistor replacing TJ.
  Here should be noted
that the estimation of Joule heat in the sample and in the tip
due to their resistances, which are less than 1~$\Omega$, yields
the value  less than 10$^{-14}$~W.
    Taking into account that the estimated resolution of our experiment
    is $\approx$5~nW,
     the observed effects cannot be associated with Joule heat
generation due to non-zero resistances of TJ electrodes.

 The results represent a direct calorimetric verification
of the model of energy dissipation in TJ as presented in
\cite{Hebard76} that can be summarized as follows.
   Considering metal P and metal N
as the positive and negative TJ electrode, respectively, separated
by a vacuum tunnel barrier as shown in Fig. 3,
 the  net tunnel current $I$
 can be expressed in the form
\begin{equation}
I = e \int_{-\infty}^{+\infty}  n(E)dE.
\end{equation}
Here $n(E)$ is the resulting number of tunneled electrons
per one second within the energy range $\langle E, E+dE \rangle$
(see Ref. \cite{NetCurr})
where $E$ are energies of tunneled electrons counted
from the Fermi level of metal P.
    As can be deduced from Fig. 3,
the process of elastic electron tunneling at energy $E$
is  accompanied by inelastic electron process(es)
 as the electron after passing the barrier appears
in the excited state with excess
energy $E$  above the Fermi level of metal P
    and therefore it has to be brought into the equilibrium state
via inelastic collisions
generating non-equilibrium quasiparticles, e.g. phonons,
with the total energy $E$.
    Analogously, as the  empty state (non-equilibrium hole)
 has been created  below
the Fermi energy  in the metal N
one has to be filled
by the electron from higher energy level.
 This is accompanied by inelastic electron process(es)
 in metal N with the total energy $(eV - E)$.
    Thus, based on the Eq. (1) and considering
that the energy of all inelastic electron processes in electrodes
will be as result converted to the heat, one can write
\begin{eqnarray}
P_{pos} = \int_{-\infty }^{+\infty }E n(E)dE\\
P_{neg} = \int_{-\infty }^{+\infty }(eV-E) n(E)dE,
\end{eqnarray}
so that, for the power dissipated in the whole TJ we have
\begin{equation}
P_{pos} + P_{neg} =  V I.
\end{equation}

As can be seen from Fig. 2 and its inset, experimental data are in
excellent quantitative correspondence with Eq. 4
    yielding the proof that CT experiment measures
    the energy dissipated in TJ electrodes in absolute units.
   The profile of $P$-$V$
curves from Fig. 2. can be explained using Eqs. (2) and (3)
as follows.
    At sufficiently low temperatures ($k_BT<<eV$)
for $E$ out of the range $\langle 0, eV \rangle$ the
$n(E)$ can be regarded as zero,
while in the range $\langle 0, eV \rangle$
it can be  approximately expressed by
$n(E) = \rho_{N}(E-eV)\tau(E, eV, z, \phi)  \rho_{P}(E)$
($\rho_P$ and $\rho_N$ are the densities of states of
the positive and negative electrode, respectively, $\phi$ is the
output work of electrodes,
$\tau$ is the transmission barrier probability).
      In the low-voltage limit ($|eV| << \phi$),
assuming normal-metal electrodes, $n(E)$
 can be considered as constant in the range $\langle 0, eV \rangle$
yielding $P_{pos} = P_{neg} = VI/2$.
    At higher bias voltage the exponential behavior of $n(E)$ has to be
taken into account,
as the energy dependence
of $n(E)$ in Eqs. (2) and (3) will be preferably
governed by $\tau(E, eV,z,\phi )$
which is an exponential function of $E$.
    As $V$ increases
the states with energy slightly lower than
$E=eV$ give ever more important
contribution to the tunnel current.
    For $P_{pos}$ the contribution
from these states  is enhanced
due to the multiplication of $n(E)$ by $E$,
 while for $P_{neg}$
the situation is just opposite, due to the term $(eV - E)$.
    This explains the steeper growth of $P_{pos}$
with comparison to $P_{neg}$ with increasing  $V$
and the tendency of $P_{neg}$ to saturate at highest $V$,
where the dominating contribution to the tunnel current
originates just from the states
from the region close below $E=eV$.

    The qualitative difference between $P$-$V$ and $P$-$I$ characteristics
explains the fact that while a change of tunnel current at
constant bias voltage changes the mutual distance between TJ
electrodes and therefore governs properties of the tunnel barrier
characterized by $\tau$, the change of bias voltage, in addition,
influences the energy scheme of TJ (and defines the
energy window for inelastic electron processes).

    As follows from discussion above,
    the unique property of CT studies is
that in contrast to  $I$-$V$ characteristic studies, where the
tunnel current is measured as a common property of  both TJ
electrodes, studies of $P$-$V$ characteristics  yield the
information about energy processes for each of TJ electrodes separately.
    Due to this, CT experiment is sensitive  to all kinds
of physical processes causing an energy transfer from/between TJ electrodes,
like  light emission caused by tunnel current \cite{Hoffmann}
or selective extraction of ``hot" electrons
from normal-metal to superconductor \cite{Nahum94}.
    However, if there is no transfer energy
from the tunnel structure, the coupling between $I$-$V$ and
$P$-$V$ dependencies  via Eqs. (1), (2) and (3) implicates the
equivalence of both approaches.
    Therefore we claim that {\em
careful} studies of $P$-$V$ characteristics can be used
for derivation of energy-spectroscopic information analogously like
``classical" tunneling spectroscopy based on measurements of
$dI/dV$ or
$d^2I/dV^2$ \cite{Wolf85}.
    [We note that at sufficiently low temperatures
    ($k_{B}T<<eV$) the equivalence
$dI/dV = (1/V)dP_{pos}/dV$ takes place, as we showed elsewhere
\cite{BatBat}.]

  Presented results clearly
indicate a different nature of charge injection in PCs and TJs.
    While in ballistic PCs the energy
gained due to the  acceleration in the field region of the contact
is expected to dissipate only  in the electrode into which the
electron is injected, in TJs the heat generation
 in {\em both} electrodes takes place.
    For the illustration, following the energy diagram in Fig. 3,
the charge injection in PCs can be
imagined as a ``resonant tunneling" directly from Fermi level
of metal N into the state with energy $eV$ above Fermi level
of metal P so that the whole energy $eV$ is
dissipated in the metal P only.
    If we characterize the asymmetry of  heat production
by $A=(P_{pos} - P_{neg})/(P_{pos} + P_{neg})$
then for the ballistic PC limit $A_{PC}=1$,
while for the case of the normal-metal --
vacuum -- normal-metal TJ
 in the low-voltage limit $A_{TJ}=0$.
    For the contacts with the ``ballistic PC --TJ crossover behavior"
(e.g. like the  crossover from metallic to TJ behavior for
normal-metal -- superconducting microconstriction contacts
introduced by Blonder, Tinkham and Klapwijk  \cite{BTK}) the value
of $A$ should lie between $A_{PC}$ and  $A_{TJ}$.
    This offers an experimental possibility for
 the determination of mutual contribution of {\em tunneled} and
 {\em ballistically injected} carriers in ``crossover" contacts.
    Of course, such type of experiments requires a non-vacuum
tunnel barrier and a consequent consideration of the thermal
coupling between the tip and the sample.

    From the point of view of possible applications of CT
    experiment
(namely in the field of design and development of electronic
devices) we would like to emphasize that a TJ represents a
source of thermal gradients,
 especially at high bias voltage and ``non-constant" behavior of
 $n(E)$.
     As well,  non-equilibrium charge carrier distribution
due to tunneling and consequent generation of
quasiparticles  (e.g. phonons), which can be effectively studied
by CT experiment,
 play fundamental role in many
applications, especially at very low temperatures
and in a small electronic devices
where thermalization of the electron system
 presents a  fundamental problem \cite{Meshke04}.
     For instance, electrical conduction in impurity bands can be
realized via phonon-assisted tunneling between localized states
and therefore generated non-equilibrium phonons
 can be a reason  for additional channel of
electrical conduction.
    Analogously, the spin relaxation rate due to phonon scattering,
which plays an important role in spintronic applications \cite{Zutic04},
 can be influenced by non-equilibrium phonons generated
due to the
 tunneling/injection of spin polarized charge carriers.
    Thus the heat generation/sinking accompanying the
charge tunneling is a rather complex task which can not be
sufficiently described only by the total dissipated power and
corresponding overheating and/or generation of thermal gradients,
as generated quasiparticles, whose energy window is defined by
applied bias voltage, can drive physical processes in each of TJ
electrodes.
    We believe that the CT experiment represents a powerful tool
suitable for studies of energy processes accompanying a charge
injection that according to its nature and its capability to be
extended to a scanning probe microscopy technique will allow to
answer many open questions important for applications in
electronics (including nanoelectronics).

In summary, the principle of novel calorimetric tunneling
experiment has been proposed. 
	Using this experiment we have
measured the heat generated in normal-metal -- vacuum --
normal-metal TJ.
    Results of the measurements show that the mutual
redistribution of the power dissipated in TJ elecrodes is
symmetrical in the low voltage limit, however, it shows a marked
asymmetry at higher bias voltage.
    Moreover, they indicate that the generated heat
represents the energy of non-equilibrium quasiparticles coming
from inelastic electron processes which accompany the process of
electron tunneling.
   According to its nature the CT experiment is sensitive
to all kinds of physical processes causing an energy transfer
from/between TJ electrodes and it can be extended to the space resolved
     techniques by means of scanning probe microscopy.
    It can be perspectively utilized for energy spectroscopy
of electrically conductive solids, studies of thermoelectric
effects in tunnel structures, as well as in development of TJ
containing devices those functionality is influenced by
temperature gradients or by generation of non-equilibrium charge
carriers and/or non-equilibrium quasiparticles, like electronic microrefrigerators
based on a NIS (or similar) tunnel structures or Coulomb blockade
thermometers.

  We thank Prof. M. Mei$\ss$ner for the opportunity to
perform our first experimental attempts to measure heat generation in VBTJs
in the Heat Capacity Laboratory at Hahn-Meitner-Institut Berlin
  and for
his useful discussions and recommendations on the calorimeter design.
   This work was supported by Centre of Low Temperature Physics
 operated as the Centre of Excellence of the Slovak Academy
of Sciences under Contract No. I/2/2003
and by the VEGA grant No. 2/4050/04.

\newpage
Figure Captions

\vspace{0.5cm}
Fig. $1.$
Schematic depiction of the CT experimental setup.

\vspace{0.5cm}
Fig. $2.$ 
 Dependence of the heat power
 on the bias voltage for positively (squares),
 negatively (triangles) biased sample,
 and  of the total power $P_{tot}= P_{pos} + P_{neg}$
(circles).
    The dotted line represents the power obtained using
the classical formula $P_{Joule}=VI$.
    Inset:
  Dependence of normalized heat power $P/P_{tot}(\hbox{100~nA})$
 on the tunnel current  for positively (squares)
and negatively (triangles) biased sample measured
at $V=\pm$2.5~V.
Normalized total power (circles)
is compared with the one calculated using
formula $VI/(V\times \hbox{100~nA})$ (dotted line).
For illustration, normalized linear fits (by means of regression
formula $y = bx$) of data
taken at $V=\pm$1.0~V
 for positively (dot-dashed line) and negatively (dashed line)
biased samples are plotted as well.

\vspace{0.5cm}
Fig. $3.$  
The energy scheme of normal-metal -- insulator --
normal-metal TJ  at external bias voltage $V$. Elastic electron
tunneling process at the energy $E$, marked by an arrow, is
followed by inelastic processes accompanied by emission of
quasiparticles with total energy $E$ and $eV-E$ in metal P and
metal N, respectively. The maximum energy of generated
quasiparticles  is $\hbar \omega_{0} = eV$.


\begin{thebibliography}{99}
\bibitem{NetCurr}
We consider  $n(E) = n_{1}(E) - n_{2}(E)$, where $n_{1}(E)$ and
$n_{2}(E)$ are numbers of tunneled electrons per one second within
the energy interval $\langle E, E+dE \rangle$ from metal N to
metal P and from metal P to metal N, respectively.
\end{thebibliography}
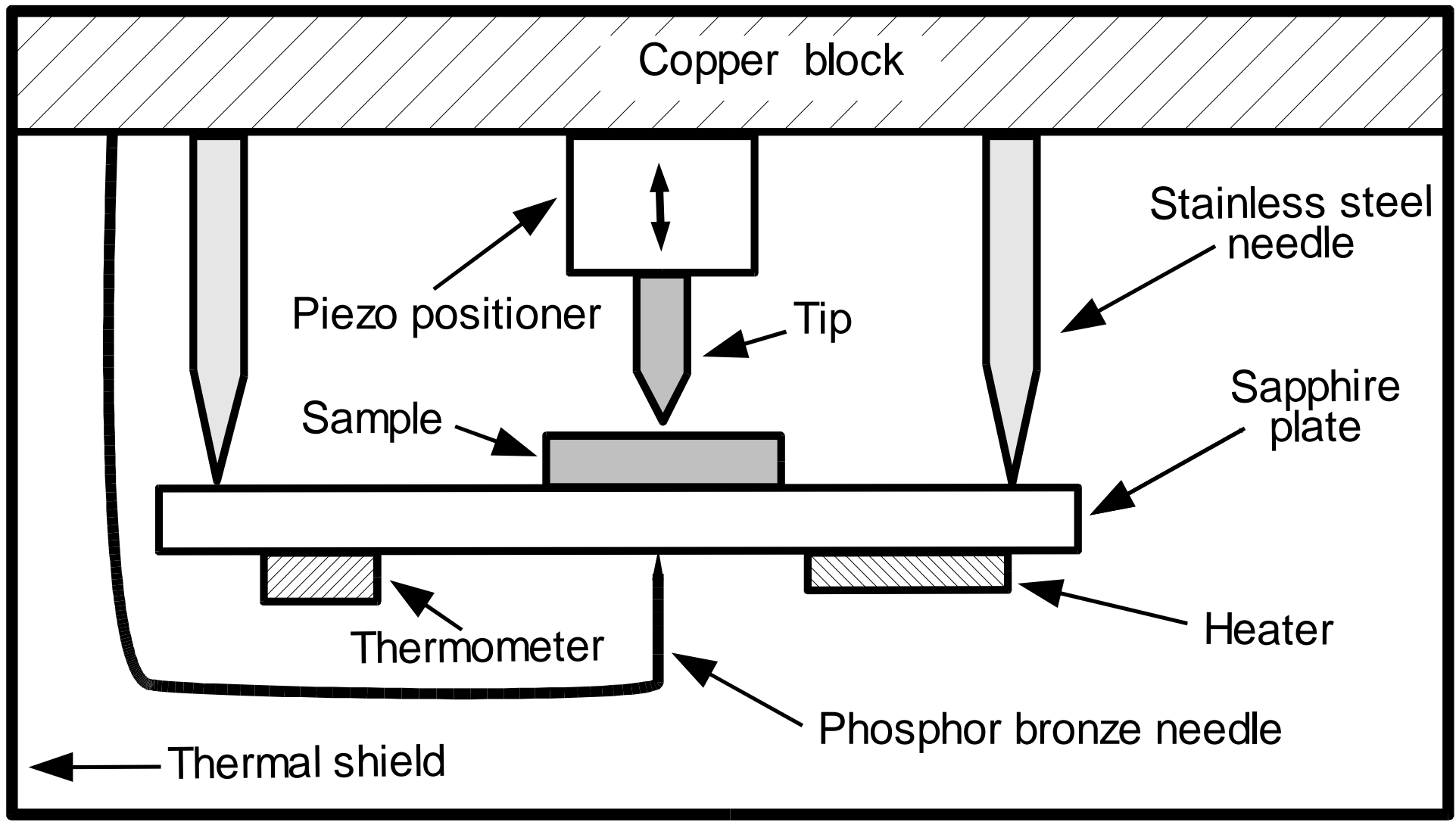
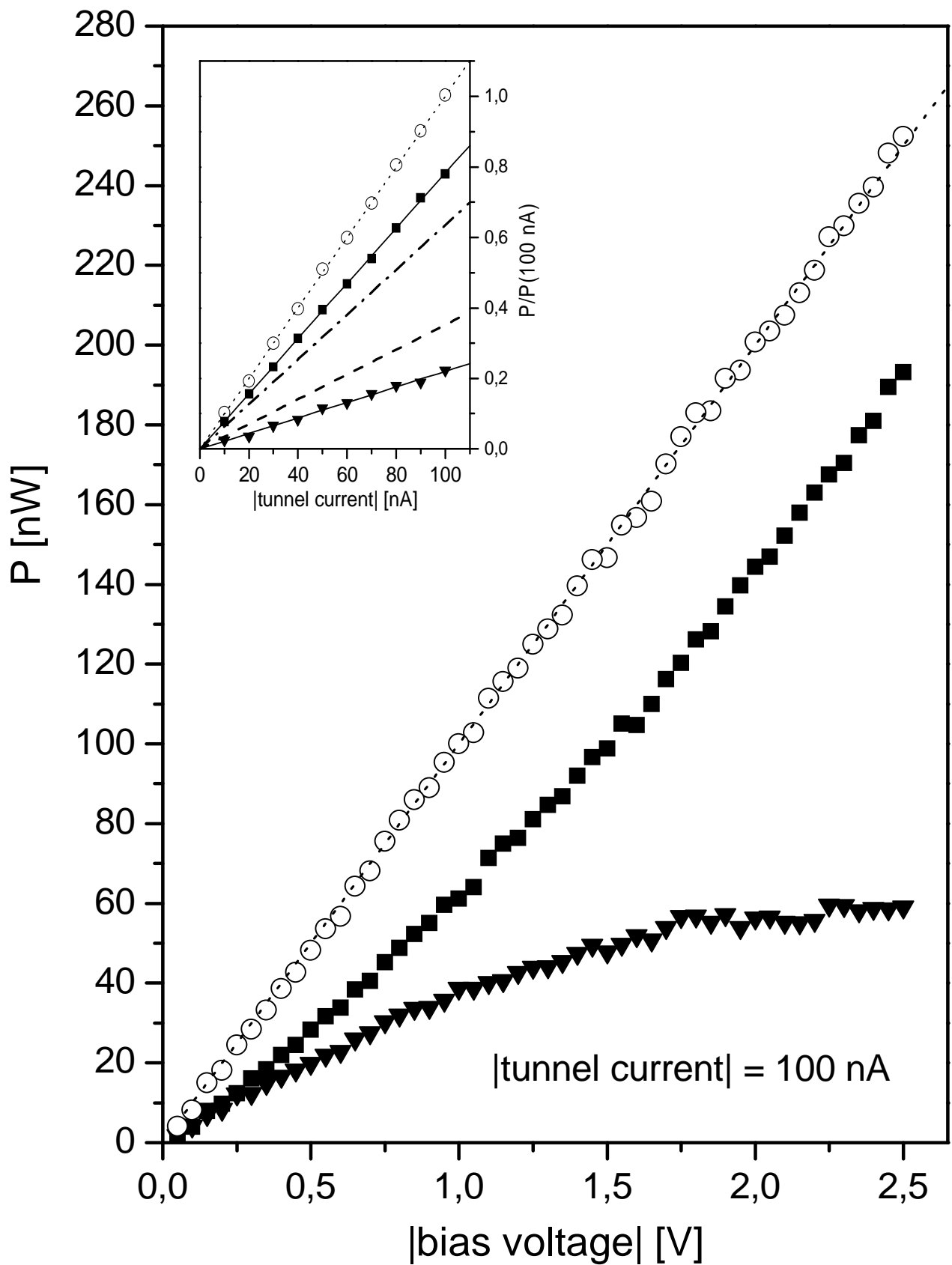
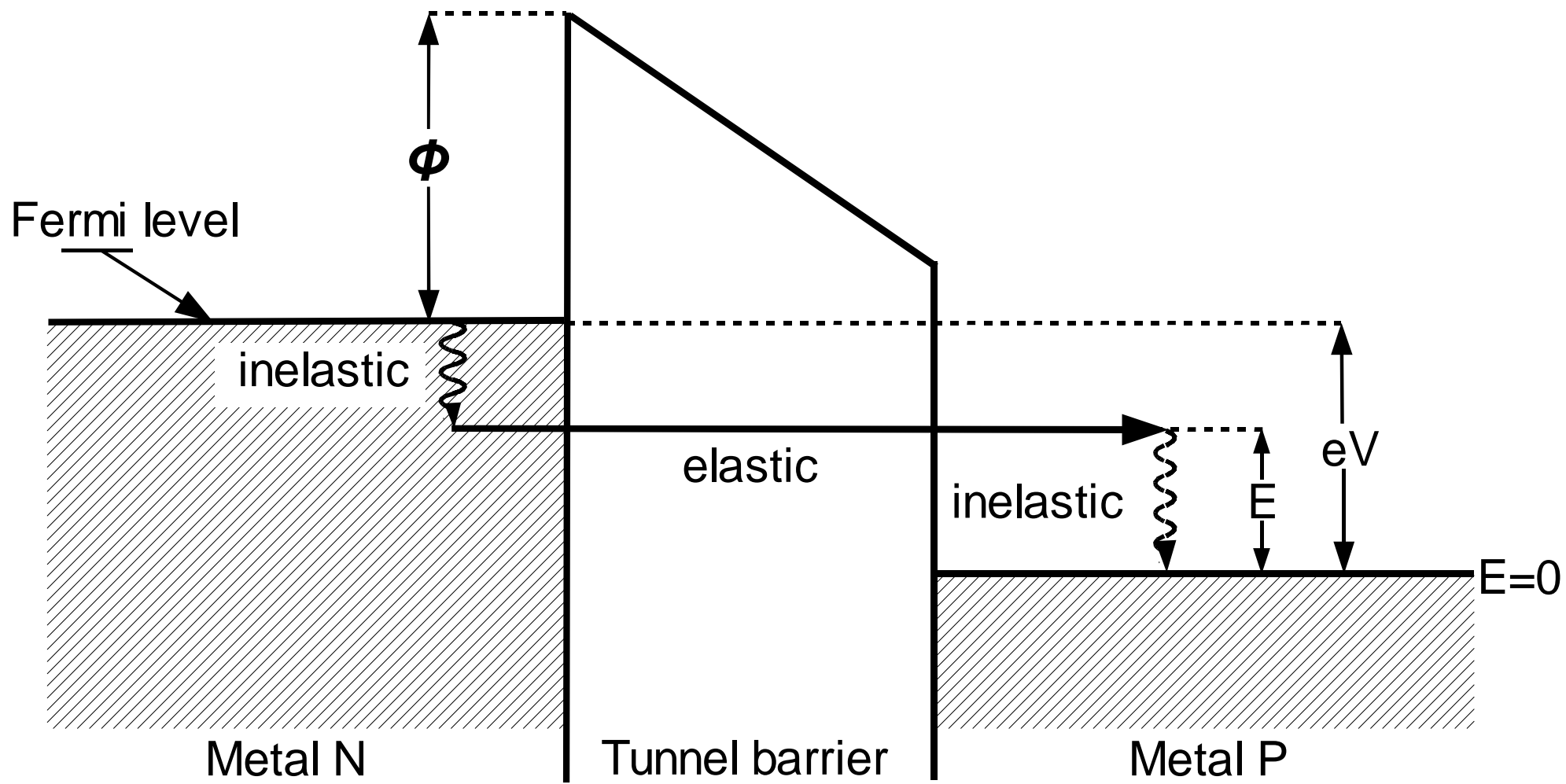
\end{document}